\documentclass[a4paper,11pt]{article}
\usepackage[a4paper,hmargin=1.1in,vmargin=1.1in]{geometry}
\usepackage[english]{babel}
\usepackage{graphicx}
\usepackage{amscd}
\usepackage{amsfonts}
\usepackage{latexsym}
\usepackage{amsthm}
\usepackage{amssymb,amsmath}
\usepackage[all]{xy}
\usepackage{bbm}
\usepackage{enumerate}
\usepackage{hyperref}

\newtheorem{Definition}{\sc Definition}[section]
\newtheorem{Theorem}[Definition]{\sc Theorem}
\newtheorem{Lemma}[Definition]{\sc Lemma}
\newtheorem{Proposition}[Definition]{\sc Proposition}
\newtheorem{Corollary}[Definition]{\sc Corollary}

\theoremstyle{definition}
\newtheorem{Remark}[Definition]{\sc Remark}

\newcommand{\BIGOP}[1]{\mathop{\mathchoice

{\raise-0.22em\hbox{\huge $#1$}}

{\raise-0.05em\hbox{\LARGE $#1$}}{\hbox{\large $#1$}}{#1}}}

\newenvironment{Proof}[1][]{\par \sc Proof{#1}. \rm\ignorespaces}{\hspace*{\fill}$\triangle$\vspace{1ex}\ignorespacesafterend}

\newcommand{\C}{\mathbb{C}}

\newcommand{\one}{\ensuremath{\mathbf{1}}}

\newcommand{\supp}{\operatorname{supp}}

\newcommand{\GL}{\operatorname{GL}}

\newcommand{\St}{\operatorname{St}}
\newcommand{\wt}{\operatorname{wt}}

\newcommand{\vspan}[1]{{\langle{#1}\rangle}}

\def\scalprod#1#2{({#1}\, |\, {#2})}

\title{Critical pairs for the Product Singleton Bound}
\author{Diego Mirandola\thanks{CWI Amsterdam and Mathematical Institute, Leiden University, The Netherlands, and Institut de Math\'ematiques de Bordeaux, Universit\'e de Bordeaux, France. Email: \texttt{diego@cwi.nl}.}~\ and Gilles Z\'emor\thanks{Institut de Math\'ematiques de Bordeaux, Universit\'e de Bordeaux, France. Email: \texttt{zemor@math.u-bordeaux.fr}.}}
\date{\today}

\begin{document}
\maketitle{\let\thefootnote\relax\footnotetext{This paper was presented at the 2015 Ninth International Workshop on Coding and Cryptography and at the 2015 Information Theory and Applications Workshop. The second author is thankful for financial support from IdEx Bordeaux - CPU (ANR-10-IDEX-03-02).}}

\begin{abstract}
We characterize Product-MDS pairs of linear codes, i.e.\
pairs of codes $C,D$ whose product under coordinatewise multiplication has maximum possible minimum distance as a function 
of the code length and the dimensions $\dim C, \dim D$. We prove in particular, for $C=D$, that if the square of the code $C$ has minimum distance at least~$2$,
and $(C,C)$ is a Product-MDS pair, then
either $C$ is a generalized Reed-Solomon code, or $C$ is a direct sum of self-dual codes. In passing we establish coding-theory analogues of classical
theorems of additive combinatorics.
\end{abstract}

\section{Introduction}
Let $F$ be a finite field. Given vectors $x=(x_1,\dots,x_n)$, $y=(y_1,\dots,y_n)$ of $F^n$, let us denote by $xy$ the componentwise product of $x$ and $y$,
$$xy=(x_1y_1,\dots,x_ny_n).$$
Given two linear codes $C,D\subseteq F^n$, let us denote by $CD$ the $F$-linear
subspace of $F^n$ generated by all products $xy$, $x\in C$, $y\in D$. This product, sometimes called the Schur product, has usually been denoted by $C*D$, but we wish to lighten notation. Likewise we shall denote the Schur square
(henceforth square) of a code $C$ by $C^2$: context should prevent confusion with cartesian products.

Products of codes turn up in a variety of situations, such as algebraic error correction, secret sharing and multiparty computation,
algebraic complexity theory, lattice constructions, and lately cryptanalysis.
For an exhaustive discussion on the connection between code products and these different topics, the reader is referred to~\cite{CCMZ15}.
A number of efforts have gone into describing the code-theoretic structure of code products, see~\cite[Chapter~12]{mpc_book} and~\cite{Randriam_draft} for an extensive review of the current state of the art.
In particular, \cite{Randriam_draft} collects several technical results which will be cited explicitly in Section~\ref{sec:preliminaries} and used later in this paper.
Among these, the following bound on the minimum distance of products, which was first proved in~\cite{Randriam13_psb}, appears.

\begin{Theorem}[Product Singleton Bound~\cite{Randriam13_psb}]\label{thm:psb_original}
 Let $C,D\subseteq F^n$ be linear codes. Then
 \begin{equation}
   \label{eq:PSB}
   d_{\min}(CD)\leq \max\{1,n-(\dim C+\dim D)+2\}.
 \end{equation}
\end{Theorem}

A slightly stronger version of Theorem~\ref{thm:psb_original} is actually proved in \cite{Randriam13_psb}, as is a version involving the product of more than two codes, but the above statement is really what motivates
our discussion. We shall call the upper bound \eqref{eq:PSB} the {\em Product Singleton Bound}, that can be thought of as a generalization of the classical
Singleton Bound. Indeed, the classical Singleton Bound 
  for a single code $C$ is recovered
by taking the code $D$ in Theorem~\ref{thm:psb_original} to be of dimension $1$ and minimum distance $n$.



Our goal in this paper is to characterize pairs $(C,D)$ of codes that
achieve equality in~\eqref{eq:PSB}. We make the remark that if $d_{\min}(CD)$
is allowed to be equal to $1$, then pairs achieving equality in \eqref{eq:PSB}
can be almost anything, since typical pairs of codes will have a product
equal to the whole space $F^n$. For a study of this phenomenon see \cite{CCMZ15, Randriam15}. So we shall disregard the situation when $d_{\min}(CD)=1$ and call $(C,D)$
a {\em Product-MDS (PMDS)} pair if it achieves equality in~\eqref{eq:PSB}
and $d_{\min}(CD)\geq 2$.

As mentioned above, a PMDS pair can consist of an ordinary MDS code and
a code of dimension $1$. It is a natural question to ask what other PMDS pairs
exist. It turns out that there is a surprisingly complete answer to this
question. We shall show in particular that if $(C,D)$ is a PMDS pair
such that $\dim C\geq 2$, $\dim D\geq 2$, and $d_{\min}(CD)\geq 3$, then
$C$ and $D$ can only be Reed-Solomon codes. 
By this we mean Reed-Solomon
code in the widest sense, i.e.\ generalized, 
possibly extended or doubly extended in the terminology of~\cite{MWS}, or Cauchy codes as in~\cite{Dur}.
PMDS pairs with $d_{\min}(CD)=2$ will also be described quite precisely.
To be more specific, in the symmetric case $C=D$ we shall prove:

\begin{Theorem}\label{thm:(C,C)}
 If $(C,C)$ is a PMDS pair, then $C$ is either a Reed-Solomon code or a direct sum of self-dual codes.
\end{Theorem}

Self-duality in the above statement should be understood to be relative to a non-degenerate bilinear form which is not necessarily the standard inner product.

To establish these results we shall import methods from additive combinatorics
and establish coding-theoretic analogues of the classical theorems of Kneser
\cite{Kneser} 
and Vosper \cite{Vosper}. For background on and proofs of Kneser and Vosper's Theorems we refer to \cite{TaoVu}. Kneser's Theorem implies in particular that
if $A,B$ are subsets of an abelian group such that 
$$|A+B|< |A|+|B|-1$$
then $A+B$ must be periodic, i.e.\ there exists a non-zero element $g$ of the
abelian group that stabilizes $A+B$ so that we have $A+B+g=A+B$.
Our coding-theoretic variant of Kneser's Theorem will imply that
if $C$ and $D$ are two codes such that
$$\dim CD < \dim C + \dim D -1,$$
then the code $C$ is necessarily the direct sum of two non-zero codes,
which is equivalent to the existence of a non-constant vector $x$ of $F^n$ such
that $xCD = CD$.

Vosper's Theorem is a characterization of pairs of subsets $A,B$ of the
integers modulo a prime $p$ with the property that $|A+B|=|A|+|B|-1$. It
states that, excluding some degenerate cases, $A,B$ must be arithmetic
progressions with the same difference. We make the remark that if a code $C$ has a generator matrix with rows $g,g\alpha,\dots,g\alpha^{k-1}$,
i.e.\ has a basis of elements in ``geometric'' progression then, provided $g$ is of weight $n$ and $\alpha$ has distinct coordinates, $C$ must be
a Reed-Solomon code. This is why a code-theoretic version of Vosper's Theorem
forces the appearance of Reed-Solomon codes. There will be some twists
to the analogy however that we shall discuss later in the paper.




The paper is organized as follows. Section~\ref{sec:preliminaries} sets up
notation and preliminary results and states our primary contribution, Theorem~\ref{thm:main}.
Section~\ref{sec:kneser} states and proves the coding-theory equivalent
of Kneser's Theorem.
Section~\ref{sec:vosper} is dedicated to a coding-theory version of Vosper's Theorem. Section~\ref{sec:main} shows how to recover a version of Randriambololona's Product Singleton Bound as a straightforward consequence of Kneser's Theorem
and goes on to derive the proof of Theorem~\ref{thm:main}.
Section~\ref{sec:comments} concludes with some comments.


\section{Preliminaries and statement of the main Theorem}\label{sec:preliminaries}
Throughout the paper $F$ will denote a finite field. We shall need,
in a couple of occasions, to deal with fields that may be infinite in
which case we will use the notation $K$. 

Given a vector $x\in K^n$, we denote by $\supp x$ its support and by $\wt(x)$ its weight. The support of a subvector space of $K^n$ is defined
as the union of the supports of all its vectors, and we shall say that
a vector space in $K^n$
has {\em full support} if its support is $\{1,\ldots ,n\}$.

In this paper all codes will be linear.
We will call them simply ``codes'' when the ambient space is $F^n$,
and use the terminology of vector spaces in the general setting of $K^n$.

\subsection{MDS Codes}\label{sec:mds}


Given a code $C\subseteq F^n$, we denote by $C^\bot$ its dual with respect to the standard inner product in $F^n$ and by $d_{\min}(C)$ its minimum distance. 

The classical Singleton Bound  states 
$$
\dim C+d_{\min}(C)\leq n+1.
$$
A code which attains this bound is said to be Maximum Distance Separable (MDS).
We recall the following well-known properties and
characterizations of MDS codes \cite{MWS}.

\begin{Lemma}\label{lemma:mds}
Given a code $C\subseteq F^n$, the following statements are equivalent:
 \begin{enumerate}
 \item $C$ is MDS,
 \item $C^\bot$ is MDS,
 \item any $\dim C$ columns of a generator matrix of $C$ are linearly independent,
 \item for any coordinate set $I$ of size $n+1-\dim C$, there is a word of $C$ whose support equals~$I$.
 \end{enumerate}
\end{Lemma}

The following property is somewhat less standard.

\begin{Lemma}\label{lemma:mds_equivalent}
Let $C\subseteq F^n$ be a code. It is MDS if and only if any systematic generator matrix of $C$ has all its rows of weight $n+1-\dim C$.
\end{Lemma}
%

\begin{Proof}
It is clear that if $C$ is MDS the property must hold.
The converse implication is an immediate consequence of the following claim: if $C\subseteq F^n$ is any code and $x\in C$ is a codeword of minimal weight, then there is a systematic generator matrix of $C$ whose first row is $x$.

We now prove this claim.
Renumbering the coordinates, we may assume that $\supp x=\{1,2+n-\wt(x),\dots,n\}$ and that
$$
x=(1,0,\dots,0,*,\dots,*),
$$
where the stars denote non-zero entries. 
Let $\{x_1=x,x_2,\dots,x_k\}$ be an $F$-basis of $C$ containing $x$, where $k:=\dim C$, and let $G$ be the generator matrix of $C$ whose rows are the $x_i$'s.
If $G$ can be made systematic in the first $1+n-\wt(x)$ positions then we are done.
Otherwise, we obtain a contradiction as follows.
We have that $\wt(x)>1$ and the rank of the matrix $G$ restricted to its first $1+n-\wt(x)$ columns
is $<k$. There exists therefore a linear combination
$$
\tilde x=\sum_{i=2}^k\alpha_ix_i,
$$
with $\alpha_2,\dots,\alpha_k\in F$, which has zeros in positions $\{2,\dots,1+n-\wt(x)\}$, but with $\tilde x\not=0$.
Now a suitable combination of $x$ and $\tilde x$ yields a non-zero word of weight smaller than $\wt(x)$, contradicting the minimality of $\wt(x)$.
\end{Proof}

Among MDS codes, Reed-Solomon codes, in the widest possible sense, will be prominent. 
A Reed-Solomon code of length $n$ and dimension $k$ is a code of the form
$$
\{(g_1f(\alpha_1),\dots,g_nf(\alpha_n)):f\in F[X]_{<k}\},
$$
where $F[X]_{<k}$ denotes the space of polynomials of degree less than $k$,
$g_1,\dots,g_n$ are non-zero elements of $F$,
and  $\alpha_1,\dots,\alpha_n$ are pairwise disjoint and
belong to $F\cup\{\infty\}$, with the convention that for any $f\in F[X]_{<k}$, $f(\infty)$ equals the coefficient of $X^{k-1}$ in $f$. We shall
call $(\alpha_1,\ldots ,\alpha_n)$ an {\em evaluation-point sequence}
for the Reed-Solomon code. 


This code family includes the codes called generalized, extended,
and doubly-extended Reed-Solomon codes. From the geometric
point of view, they may be thought of as the projective version of Reed-Solomon codes. In~\cite{Dur} they are named ``Cauchy codes''
and have also been called ``Cauchy Reed-Solomon codes''. We shall simply refer
to them as ``Reed-Solomon codes''.

\begin{Remark}
%
If $C$ and $D$ are two Reed-Solomon codes with a common evaluation-point
sequence $\alpha$, then the product $CD$ is also Reed-Solomon with evaluation-point sequence
$\alpha$ and we have $\dim CD=\min\{n,\dim C+\dim D-1\}$. Theorem~\ref{thm:mdsbound}
below implies that $\min\{n,\dim C+\dim D-1\}$ is the minimum possible
dimension of the product of MDS codes.
\end{Remark}

\subsection{Code products}
For an arbitrary field $K$,
the space $K^n$ is, with the coordinatewise product, 
a commutative unitary $K$-algebra.
Its unit element is the all-one vector, denoted by $\one$.
The multiplicative group of its invertible elements is 
$(K^n)^{\times}=(K^{\times})^n$, meaning that $x\in K^n$ is invertible if and only if all entries of $x$ are non-zero. Given $x\in(K^n)^{\times}$, 
its inverse is denoted by $x^{-1}$. The following simple observation will be freely  used later.

\begin{Lemma}\label{lemma:2.4}
Let $x\in (K^n)^{\times}$.
For any vector space $V\subseteq K^n$, we have $\dim V=\dim xV$.
\end{Lemma}


The two results below relate the dimension of the product of two codes with the MDS property.
The first one is taken from~\cite[\S 3.5]{Randriam_draft}.

\begin{Theorem}\label{thm:mdsbound}
Let $C,D\subseteq F^n$ be full-support codes. If (at least) one of them is MDS, then
$$
\dim CD\geq \min\{n,\dim C+\dim D-1\}.
$$
\end{Theorem}

We also observe the following.

\begin{Lemma}\label{lemma:mds_product}
Let $C,D\subseteq F^n$ be MDS codes such that 
$$
\dim CD=\dim C+\dim D-1.
$$
Then $CD$ is MDS.
\end{Lemma}
\begin{Proof}
By Lemma~\ref{lemma:mds}, it suffices to show that for any choice of 
$I\subseteq\{1,\dots,n\}$ with $|I|=d^*:=n+1-\dim CD$ there exist $x\in C,y\in D$ with $\supp xy=I$.
Without loss of generality, assume that $I=\{1,\dots,d^*\}$.
As $C$ and $D$ are both MDS, there exist $x\in C$ and $y\in D$ such that $\supp x=I\cup\{d^*+1,\dots,d_C\}$ and $\supp y=I\cup\{n-(d_D-d^*)+1,\dots,n\}$, where $d_C$ and $d_D$ denote the minimum distance of $C$ and $D$ respectively.
One checks that $d_C= n-(d_D-d^*)$, hence indeed $\supp xy=\supp x\cap\supp y=I$. 
\end{Proof}

\subsection{Stabilizer algebras and decomposable codes}

Lemma~\ref{lemma:algebra} below classifies all subalgebras of $K^n$.
For all $i=1,\dots,n$, let $e_i$ denote the $i$-th unit vector in $K^n$. 
We call a vector of the form $\sum_{i\in I}e_i$ for some $I\subseteq\{1,\dots,n\}$ a {\em projector}.
In particular, $\one$ is the projector with support $\{1,\dots,n\}$. A family of projectors is {\em disjoint} if the projectors have pairwise disjoint supports.

\begin{Lemma}\label{lemma:algebra}
Any $K$-subalgebra of $K^n$ admits a $K$-basis of disjoint projectors.
\end{Lemma}
\begin{Proof}
Let $A\subseteq K^n$ be a $K$-subalgebra.
We argue by induction on $k:=\dim A$. If $k=1$ then $A$ is generated by a
vector $x$
whose non-zero coordinates must be all equal, otherwise $x^2$ is not a $K$-multiple of $x$. 
If $k>1$, pick $x\in A,x\not=0$ of minimal support with one of its
coordinates equal to $1$, and let $\{x_1=x,\dots,x_k\}$ be a $K$-basis of $A$ containing $x$. 
Then $x$ is a projector, otherwise $x^2-x\not=0$ would have smaller support. For all $i=2,\dots,k$, 
if $\supp x_i$ and $\supp x$ intersect, say in position $j$, then we can choose $\lambda_i\in K$ so that $x_i + \lambda_i x$ has a zero in position $j$, hence $x(x_i + \lambda_i x)$ has support strictly smaller than $x$. By minimality of $\supp x$, $x(x_i + \lambda_i x)=0$, i.e.\ $x$ and $x_i + \lambda_i x$ have disjoint support.
Replacing if need be $x_i$ by $x_i+\lambda_ix$, we obtain that 
 $A$ is a direct sum $A=\vspan{x}\oplus\vspan{x_2,\dots,x_k}$ and the conclusion follows by applying the induction hypothesis to the second summand.
\end{Proof}

\begin{Remark}\label{rem:finite}
Lemma~\ref{lemma:algebra} implies in particular that the number of subalgebras of $K^n$ is finite. This fact will be useful later.
\end{Remark}

Let $V\subseteq K^n$ be a $K$-vector space. We define $\St(V):=\{x\in K^n:xV\subseteq V\}$, the {\em stabilizer} of $V$ in $K^n$.
As $V$ is linear, $\St(V)$ is a $K$-algebra, hence Lemma~\ref{lemma:algebra} applies. In particular, $\St(V)$ has a basis of vectors whose entries are all $0$'s and $1$'s.
This leads to the following lemma, that characterizes the stabilizer of a base-field extension\footnote{If $K'/K$ is a field extension, the base-field extension $V\otimes K'$, where the tensor product is taken over $K$, of $V$ is the $K'$-span of $V$.}. For a proof we refer to~\cite[Proposition 2.24]{Randriam_draft}.

\begin{Lemma}\label{lemma:fieldextension}
Let $K'/K$ be a field extension. Let $V\subseteq K^n$ be a $K$-vector space. Then
$$
\St(V\otimes K')=\St(V)\otimes K'.
$$
\end{Lemma}

Let $C\subseteq F^n$ be a code. As in the vector-space case, we can define its stabilizer and apply Lemma~\ref{lemma:algebra}, which yields an $F$-basis $\{\pi_1,\dots,\pi_h\}$ of $\St(C)$ of disjoint projectors, with $h:=\dim \St(C)$. When $h=1$ we say that $C$ has trivial stabilizer. We have the following lemma, whose proof is straightforward.


\begin{Lemma}\label{lemma:stabilizerdecomposition}
Any full-support code $C\subseteq F^n$ decomposes as $$C=\pi_1C\oplus\dots\oplus\pi_hC$$
where $\{\pi_1,\dots,\pi_h\}$ is an $F$-basis of disjoint projectors of $\St(C)$.
Moreover, each summand $\pi_iC$, viewed as a code in $F^{|\supp\pi_i|}$, has trivial stabilizer and full support.
\end{Lemma}

Facts on stabilizers, including Lemmas~\ref{lemma:fieldextension} and~\ref{lemma:stabilizerdecomposition},
can be found in~\cite[from \S 2.6 onwards]{Randriam_draft}. Following
\cite{Randriam_draft}, let us say that a code is {\em indecomposable}
if it has trivial stabilizer.

Lemma~\ref{lemma:stabilizerdecomposition} states in particular that
a full-support code has non-trivial stabilizer if and only if it decomposes as a direct sum of codes, and the dimension of the stabilizer equals the number of indecomposable components.
It follows that all MDS codes that are not equal to $F^n$ 
have trivial stabilizer.

We conclude this section with two refinements of the classical Singleton Bound, involving the dimension of  $\St(C)$ beside the usual parameters. They naturally reduce to the classical Singleton Bound when the code $C$ is indecomposable,
i.e.\ $\dim\St(C)=1$.

\begin{Lemma}\label{lemma:stabilizer}
Let $C\subseteq F^n$ be a code.
\begin{enumerate}
\item If $d_{\min}(C)>1$ then
$$
d_{\min}(C)\leq n-\dim C+1-(\dim\St(C)-1).
$$
\item If $C$ has full support then
$$
d_{\min}(C)\leq \frac{n-\dim C}{\dim\St(C)}+1.
$$
\end{enumerate}
\end{Lemma}
\begin{Proof}
We may assume that $C$ has full support, as the first claim in the general case follows immediately from the first claim in the full-support case.
Set $k:=\dim C$, $d:=d_{\min}(C)$ and $h:=\dim\St(C)$. By Lemma~\ref{lemma:stabilizerdecomposition} we have that $C$ is a direct sum $C=C_1\oplus\dots\oplus C_h$ of full-support codes.
For all $i=1,\dots,h$, let $n_i,k_i$ and $d_i$ denote the support size, the dimension and the minimum distance of $C_i$ respectively. We have $\sum_{i=1}^hn_i=n,\sum_{i=1}^hk_i=k$ and 
$$
d=\min_i\{d_i\}\leq\min_i\{n_i-k_i\}+1
$$
by the classical Singleton Bound.
In the case $d>1$ we have $n_i-k_i\geq d_i-1\geq1$ for all $i=1,\dots,h$,
hence, for all $j=1,\dots,h$,
$$
n_j-k_j= n-k-\sum_{i\neq j}(n_i-k_i)\leq n-k-(h-1).
$$
Putting everything together we have
$$
d= \min_i\{d_i\}\leq\min_i\{n_i-k_i\}+1\leq n-k+1-(h-1),
$$
which proves the first claim.
To prove the second claim, note that
$$
n-k=\sum_{i=1}^h(n_i-k_i)\geq h \min_i\{n_i-k_i\},
$$
hence $\min_i\{n_i-k_i\}\leq (n-k)/h$ and the conclusion follows.
\end{Proof}

\subsection{Main Theorem}
Our main result takes the following form.
\begin{Theorem}\label{thm:main}
Let $C,D\subseteq F^n$ be codes such that the pair $(C,D)$ is Product MDS. Then one of the following situations occurs.
\begin{enumerate}[(i)]
\item $C$ and $D$ are MDS and, if none of them has dimension $1$, they are
Reed-Solomon codes with a common evaluation-point sequence.
\item 
There is a partition of the coordinate set into non-empty subsets
$$\{1,\dots,n\}=I_1\cup\cdots\cup I_h$$
and there exist $h$ pairs  $(C_1,D_1),\ldots ,(C_h,D_h)$ of codes of $F^n$,
such that $\supp C_i=\supp D_i=I_i$, for all $i=1,\ldots ,h$, 
and such that $C$ and $D$ decompose as:
\begin{align*}
  C &=C_1\oplus\dots\oplus C_h,\\
  D &=D_1\oplus\dots\oplus D_h.
\end{align*}
Furthermore, for all $i=1,\dots,h$,
when $C_i$ and $D_i$ are identified with codes of $F^{|I_i|}$
through the natural projection on their support, we have that $C_i=(g_iD_i)^\bot$ for some $g_i\in (F^{|I_i|})^\times$.
\end{enumerate}
\end{Theorem}

\begin{Remark}
 The codes $C_i$ and $D_i$ are mutually orthogonal relative to the non-degenerate bilinear form $(x,y) \mapsto \scalprod{x}{g_iy}=\scalprod{g_ix}{y}$, where $\scalprod{\cdot}{\cdot}$
denotes the standard inner product. Hence the wording of Theorem~\ref{thm:(C,C)}
in the case $C=D$.
\end{Remark}

\section{Kneser's Theorem}\label{sec:kneser}

Kneser's Addition Theorem below involves the stabilizer $\St(X)=\{g\in G: g+X=X\}$ of a subset $X$ of an abelian group $G$. The (Minkowski) sum $A+B$
of two subsets $A,B$ of $G$ is defined as the set of sums $a+b$ when $a$ and $b$ range over $A$ and $B$ respectively.

\begin{Theorem}[Kneser~\cite{Kneser}]\label{thm:kneser_original}
Let $G$ be an abelian group. Let $A,B\subseteq G$ be non-empty, finite subsets.
Then $$|A+B|\geq |A|+|B|-|\St(A+B)|.$$
\end{Theorem}

Kneser's original Theorem was transposed to the extension
field setting by Hou, Leung and Xiang in~\cite{HLX02}. Let $L/K$ be 
a field extension. For $K$-linear subspaces $S,T\subseteq L$, we may
consider the product of subspaces $ST$ defined as the $K$-linear span
of the set of elements of the form $st, s\in S, t\in T$. Hou et al.'s Theorem
is concerned with the structure of pairs of subspaces whose product has
small dimension. Again, the stabilizer of a $K$-subspace $X\subseteq L$
is involved and is defined in the expected way $\St(X)=\{z\in L: zX\subseteq X\}$.

\begin{Theorem}[Generalized Kneser Theorem~\cite{HLX02}]\label{thm:hlx}
Let $L/K$ be a separable field extension. Let $S,T\subseteq L$ be non-zero, finite-dimensional $K$-vector spaces. Then
$$
\dim ST\geq \dim S+\dim T-\dim \St(ST).
$$
\end{Theorem}

Remarkably, Kneser's original Theorem for groups can be recovered easily from Hou et al.'s version.

We will now proceed to show that there is a variant of Kneser's Theorem for the algebra induced by coordinatewise multiplication.

\begin{Theorem}\label{thm:kneser}
Let $S,T\subseteq K^n$ be non-zero $K$-vector spaces. 
Then 
$$
\dim ST\geq\dim S+\dim T-\dim \St(ST).
$$
\end{Theorem}

\begin{Remark}
 The products $ST$ in Theorems~\ref{thm:hlx} and~\ref{thm:kneser} are
 in different algebras. The statement of Theorem~\ref{thm:hlx} is the
 only instance of the paper
 where the product $ST$ does not refer to a coordinatewise
 product.
\end{Remark}

\begin{Remark}\label{remark:support}
Assuming that Theorem~\ref{thm:kneser} holds in the case of full-support $S$ and $T$, the general case can be derived as follows.
Let $S_0,T_0\subseteq K^{n_0}$ be the projections of $S,T$ respectively on $\supp ST$, where $n_0:=|\supp ST|$. The spaces $S_0$ and $T_0$ both have full support, hence
$$
\dim S_0T_0\geq\dim S_0+\dim T_0-\dim \St(S_0T_0).
$$
Clearly $\dim S_0T_0=\dim ST$ and $\dim \St(ST)=\dim \St(S_0T_0)+n-n_0$.
It remains to prove that
$$
\dim S_0+\dim T_0\geq\dim S+\dim T-(n-n_0).
$$
Let $S_1,T_1\subseteq K^{n-n_0}$ be the projections of $S,T$ respectively on the complement of $\supp ST$.
Observe that $\supp S_1$ and $\supp T_1$ cannot intersect, hence $\dim S_1+\dim T_1\leq n-n_0$. Moreover $\dim S\leq \dim S_0+\dim S_1$ and $\dim T\leq \dim T_0+\dim T_1$. Putting everything together we obtain the desired inequality.
%
\end{Remark}

From here on ``Kneser's Theorem'' will refer to Theorem~\ref{thm:kneser} 
rather than to the original result.
Our proof is strongly inspired by Hou et al.'s proof of 
Theorem~\ref{thm:hlx}~\cite{HLX02}, itself drawing upon
the $e$-transform technique of additive combinatorics (see e.g.~\cite{TaoVu}).

If $V$ is a $K$-subspace of $K^n$, we use the notation $V^\times$ to
mean the subset of invertible elements of $V$.

\begin{Lemma}\label{lemma:transform}
Let $S,T\subseteq K^n$ be non-zero $K$-vector spaces. Assume that $T$ has a basis of invertible elements. Then, for all $x\in S^\times$, there exist a $K$-algebra $H_x\subseteq K^n$ and a $K$-vector space $V_x\subseteq K^n$ such that $H_xV_x=V_x$, $xT\subseteq V_x\subseteq ST$ and
$$
\dim V_x+\dim H_x\geq\dim S+\dim T.$$
\end{Lemma}
\begin{Proof}
Assume that the lemma is proved for $x=\one$. 
Then, if $S^\times$ is non-empty, for any $x\in S^\times$ we may
apply the result for the case $x=\one$ to $x^{-1}S$ and $T$. So we only need
to prove the Lemma for $\one\in S$ and $x=\one$. Analogously, we may assume that $\one\in T$.

We argue by induction on $k:=\dim S$. If $k=1$, $H:=K\one$ and $V:=T$ do the job.
So assume that $k>1$ and the result holds for smaller dimension. For each $e\in T^\times$, define
$$
S(e):=S\cap Te^{-1},\quad T(e):=T+Se.
$$
We have $S(e)T\subseteq ST$, $S(e)Se\subseteq TS$, therefore $S(e)T(e)\subseteq ST$. Furthermore, 
\begin{align*}
  \dim T(e) &= \dim T + \dim Se - \dim(T\cap Se)\\
            &= \dim T + \dim S  - \dim(Te^{-1}\cap S)
\end{align*}
by Lemma~\ref{lemma:2.4}, hence
$$\dim S(e)+\dim T(e)=\dim S+\dim T.$$
We distinguish two cases.

Assume that $S(e)=S$ for all $e\in T^\times$, 
i.e.\ $S\subseteq Te^{-1}$ for all $e\in T^\times$. Then,
since $T$ has a basis of invertible elements, we have $ST\subseteq T$.
The result then holds with 
$H$ the subalgebra generated by $S$ and $V:=T$.

Assume that there exists $e\in T^\times$ such that $S(e)\subsetneqq S$. Then $0<\dim S(e)<k$ hence the induction hypothesis applied to $S(e)$ and $T(e)$ gives an algebra $H$ and a vector space $V$ such that $HV=V$, $$T\subseteq T(e)\subseteq V\subseteq S(e)T(e)\subseteq ST$$ and
$$
\dim V+\dim H\geq \dim S(e)+\dim T(e)=\dim S+\dim T.
$$
\end{Proof}

\begin{Proof}[ of Theorem~\ref{thm:kneser}]
By Remark~\ref{remark:support} we may assume that both $S$ and $T$ have full support.
The key to the proof is the following observation. Assume that $T$ has a basis of invertible elements.
Recall that, by Lemma~\ref{lemma:transform}, for all $x\in S^\times$ there exist a $K$-algebra $H_x\subseteq K^n$ and a $K$-vector space $V_x\subseteq F^n$ such that 
\begin{eqnarray}
  H_x V_x&=&V_x\label{eq:Vx}\\ 
  xT&\subseteq& V_x\subseteq ST\label{eq:ST}\\ 
  \dim V_x+\dim H_x&\geq&\dim S+\dim T. \label{eq:H_x}
\end{eqnarray}
Set $k:=\dim S$ and assume furthermore
that there exists a $K$-basis $\{x_1,\dots,x_k\}$ of $S$ contained in $S^\times$ such that 
\begin{equation}\label{eq:H}
  H_{x_1}=\dots=H_{x_k}=:H.
\end{equation}
Then $ST=V_{x_1}+\dots+V_{x_k}$ by \eqref{eq:ST}, and therefore
$HST=ST$ by \eqref{eq:Vx}, in other words 
$H\subseteq \St(ST)$.
From \eqref{eq:H_x} it follows therefore that
$$
\dim ST+\dim \St(ST)\geq\dim V_{x_1}+\dim H\geq\dim S+\dim T,
$$
hence the conclusion.

We shall first prove the Theorem when $K$ is an infinite field, 
by showing in that case that $T$ always has a basis of invertible elements
and that there always exists a basis $\{x_1,\ldots ,x_n\}$
of invertible elements of $S$ satisfying \eqref{eq:H}.

Since $T$ has full support, it should be clear enough that
it has a basis of invertible elements for $K$ infinite.
In this case Lemma~\ref{lemma:transform} applies.
Now fix a $K$-basis $\{s_1,\dots,s_k\}$ of $S$ and define, 
for all $\alpha\in K$, $y_\alpha:=\sum_{i=1}^k\alpha^{i-1}s_i\in S$. 
For any choice of non-zero, pairwise distinct $\alpha_1,\dots,\alpha_k\in K$,
the matrix transforming $s_1,\ldots ,s_k$ into $y_{\alpha_1},\ldots ,y_{\alpha_k}$ is Vandermonde, and therefore $y_{\alpha_1},\ldots ,y_{\alpha_k}$ is
also a $K$-basis of $S$.
We now observe that the set $\{\alpha\in K:y_\alpha\in S^\times\}$ is infinite: indeed its complement in $K$ is finite, as it is a finite union of zero-sets of non-zero polynomials. That these polynomials are non-zero
is guaranteed by the full-support property of $S$. 
On the other hand, the number of subalgebras of $K^n$
 is finite by Remark~\ref{rem:finite}, in particular the number 
of subalgebras $H_x$ guaranteed by Lemma~\ref{lemma:transform} is finite.
It follows that there exist $\alpha_1,\dots,\alpha_k$ such that 
$\{x_1=y_{\alpha_1},\dots,x_k=y_{\alpha_k}\}$ is a $K$-basis of $S$ whose elements are all invertible 
and such that $H_{x_1}=\dots=H_{x_k}$.
This concludes the proof in the case $K$ infinite.

Assume now that $K$ is finite, and consider an infinite field extension $K'$ of $K$, for example the rational function field $K':=K(t)$, where $t$ is transcendental over $K$.
The infinite base-field case applies to $K'$-vector spaces. Our purpose is to draw our conclusion from this.
Define the base-field extensions 
$S':=S\otimes K',T':=T\otimes K'$, where tensor products are taken over $K$. By construction $S'$ and $T'$ are $K'$-vector spaces and we have just proved that
$$
\dim_{K'} S'T'\geq\dim_{K'} S'+\dim_{K'} T'-\dim_{K'} \St(S'T').
$$
It is clear that $S'T'=ST\otimes K'$, $\dim_{K'} S'=\dim S$, $\dim_{K'}T'=\dim T$ and $\dim_{K'} S'T'=\dim ST$, where non-indexed dimensions are taken over $K$.
Moreover $\St(S'T')=\St(ST)\otimes K'$ by Lemma~\ref{lemma:fieldextension} and the conclusion follows.
\end{Proof}

Theorem~\ref{thm:kneser} implies in particular that if $C$ and $D$ are two
codes such that $CD$ has trivial stabilizer, i.e.\ is indecomposable, then
we must have 
\begin{equation}
  \label{eq:cauchy}
  \dim CD \geq \dim C + \dim D -1.
\end{equation}
The next section studies pairs of codes $C,D$ such that $CD$ is
indecomposable and achieves equality in \eqref{eq:cauchy}.
\section{Vosper's Theorem}\label{sec:vosper}


We start by recalling Vosper's Addition Theorem.

\begin{Theorem}[Vosper~\cite{Vosper}]\label{thm:vosper_original}
Let $G$ be an abelian group of prime order $p$. Let $A,B\subseteq G$ be subsets, with $|A|,|B|\geq 2$ and $|A+B|\leq p-2$. If
$$
|A+B|=|A|+|B|-1
$$
then $A$ and $B$ are arithmetic progressions with the same difference.
\end{Theorem}



We point out that an extension-field version of Vosper's Theorem 
for finite fields was recently proved in \cite{BSZ15}. 

Since the stabilizer of a subset of a group $G$ must be a subgroup,
when $G$ is of prime order and has no proper subgroup,
Kneser's Addition Theorem~\ref{thm:kneser_original} implies that subsets $A,B$
of $G$ such that $A+B\neq G$ must satisfy 
 $$|A+B|\geq|A|+|B|-1.$$
This result is known as the Cauchy-Davenport Inequality, see~\cite{Nathanson, TaoVu}. Vosper's Theorem is therefore concerned with characterizing
pairs of sets achieving equality in the Cauchy-Davenport Inequality.

In the algebra setting, the inequality \eqref{eq:cauchy} may be thought
of as a code-product version of the Cauchy-Davenport Inequality. But contrary to
the group case, the algebra $F^n$ always has proper subalgebras (for $n>1$) 
so we cannot hope to ensure~\eqref{eq:cauchy} purely by a condition on $F^n$. 
However, we have seen that~\eqref{eq:cauchy} holds when (at least one of) 
the codes
involved is MDS (Theorem~\ref{thm:mdsbound}). The following theorem may be seen
as a version of Vosper's Theorem for MDS codes, and is the main result of
this section.

\begin{Theorem}\label{thm:vosper}
Let $C,D\subseteq F^n$ be MDS codes, with $\dim C,\dim D\geq 2$ and $\dim CD\leq n-2$. If
$$
\dim CD=\dim C+\dim D-1
$$
then $C$ and $D$ are Reed-Solomon codes with a common evaluation-point sequence.
\end{Theorem}

\begin{Remark}
The hypotheses $\dim C,\dim D\geq 2$ clearly cannot be removed.
The value $n-2$ is also best possible
in the hypothesis $\dim CD\leq n-2$, since by taking $C$ to be an arbitrary MDS
(non Reed-Solomon) code, and taking $D=C^\bot$, we will have a pair
of MDS codes such that $\dim CD = \dim C + \dim D -1 = n-1$.
\end{Remark}

We introduce the following notation for 
Vandermonde-type matrices.
Given a positive integer $k$ and $\alpha=(\alpha_1,\dots,\alpha_n)\in (F\cup\{\infty\})^n$ we denote by $V_k(\alpha)$ the $k\times n$ matrix whose $i$-th column is $(1,\alpha_i,\dots,\alpha_i^{k-1})^T$ if $\alpha_i\not=\infty$, $(0,\dots,0,1)^T$ otherwise. Note that the possible presence of this last column makes
$V_k(\alpha)$ a Vandermonde matrix in a generalized sense.
We remark that if the entries of $\alpha$ are pairwise distinct then $V_k(\alpha)$ has full rank. 
With this notation, a Reed-Solomon code of length $n$ and dimension $k$ is a code of the form $gC$, where $g\in (F^\times)^n$ (i.e.\ $g$ has no zero entries) and $C$ is generated by $V_k(\alpha)$ for some $\alpha\in (F\cup\{\infty\})^n$ with pairwise-distinct entries. 
The vector $\alpha$ is an evaluation-point sequence of $C$ (see the end
of Section~\ref{sec:mds}).

\begin{Lemma}\label{lemma:lemma4}
Let $C\subseteq F^n$ be a full-support
code with $\dim C\geq2$ and $d_{\min}(C)>1$. Assume that there exists a $2$-dimensional MDS code $A\subseteq F^n$, generated by $V_2(\alpha)$ for some $\alpha\in F^n$ with pairwise distinct entries, such that
$$
\dim AC=\dim C+1\leq n-1.
$$
Then $C$ is generated by $gV_{\dim C}(\alpha)$ for some $g\in C$. 
\end{Lemma}
\begin{Proof}
Since $\alpha$ has at most one zero coordinate, $d_{\min}(C)>1$ implies that
$\dim \alpha C=\dim C$. We therefore have
$$
\dim AC=\dim(C+\alpha C)=2\dim C-\dim(C\cap \alpha C),
$$
hence
$$
\dim(C\cap \alpha C)=\dim C-1.
$$
Moreover, $C'=C\cap\alpha C$ has support strictly larger than its dimension, otherwise it would have minimum distance $1$ and this would imply the existence of a word of weight $1$ in $C$.
We prove the lemma by induction on $k:=\dim C$.

In the case
$k=2$, pick $g'\in C\cap\alpha C$, which exists as $\dim(C\cap\alpha C)=1$, and let $g\in C$ be such that $g'=g\alpha$.
Then $g$ and $g'=g\alpha$ are linearly independent, as 
$|\supp g|\geq |\supp g'|\geq 2$ and $\alpha$ has pairwise distinct entries.
It follows that $C$ is generated by $g$ and $g\alpha$, i.e.\ by $gV_2(\alpha)$.

Now assume that $k>2$. We have
$$
k=\dim C'+1\leq \dim AC'\leq \dim\alpha C=k,
$$
where the right inequality
follows from
the inclusion $AC'= C'+\alpha C'\subseteq \alpha C$,
and the left inequality follows from
Theorem~\ref{thm:mdsbound} (recall that $A$ is MDS). Strictly speaking,
Theorem~\ref{thm:mdsbound} only applies to full-support codes and
$C'$ may have a support of cardinality $n-1$ if $\alpha$ has a
zero coordinate. But if this happens we may puncture $A$ and $C'$
by deleting this coordinate
to obtain full-support codes of the same dimension as $A$ and $C$
and still apply Theorem~\ref{thm:mdsbound}.

Since $C'\subseteq C$ we have $d_{\min}(C')\geq d_{\min}(C)>1$, and we have
just shown $\dim AC' = \dim C' + 1\leq (n-1) -1$, since $\dim C'=\dim C-1$.
Therefore the induction hypothesis applies to $C'$, possibly after puncturing
one zero coordinate to make $C'$ full support. Hence
$C'$ is generated by $g'V_{k-1}(\alpha)$ for some $g'\in C'$. 
Let $g\in C$ be such that $g'=g\alpha$.
The matrix whose rows are the elements of the set $\{g,g'=g\alpha,\dots,g'\alpha^{k-2}=g\alpha^{k-1}\}\subseteq C$ is $gV_k(\alpha)$, 
which has rank $k$ as $|\supp g|\geq|\supp C'|\geq k$.
It follows that this set is linearly independent and 
$gV_k(\alpha)$ generates $C$.
\end{Proof}



\begin{Lemma}\label{lemma:columns}
Let $C,D\subseteq F^n$ be MDS codes satisfying
$$
\dim CD=\dim C+\dim D-1.
$$
Assume that there exists an index set $I\subseteq\{1,\dots,n\}$ with $|I|\geq \dim CD$ such that the punctured
codes $C_I,D_I\subseteq F^{|I|}$ obtained by projecting $C$ and $D$ on the coordinates indexed by~$I$ are Reed-Solomon codes with a common evaluation-point sequence. 
Then $C$ and $D$ are Reed-Solomon codes with a common evaluation-point sequence.
\end{Lemma}
\begin{Proof}
Set $k:=\dim C$, $\ell:=\dim D$. Since $|I|\geq \dim CD$ we have
$|I|\geq k$ and $|I|\geq \ell$ and since $C$ and $D$ are MDS we must
have $\dim C_I=\dim C=k$, $\dim D_I=\dim D=\ell$. Note that
we may suppose $k,\ell\geq 2$, otherwise there is nothing to prove.

Reformulating the hypothesis,
there exist $g_I,g_I'\in F^{|I|}$, $\alpha_I\in(F\cup\{\infty\})^{|I|}$, where $\alpha_I$ has pairwise-distinct entries, such that $C_I$ and $D_I$ are generated by $g_IV_{k}(\alpha_I)$ and $g_I'V_{\ell}(\alpha_I)$ respectively. In other words there are unique generator matrices $G_C$ and $G_D$ of $C$ and $D$ 
whose $I$-indexed columns form $g_IV_{k}(\alpha_I)$ and $g_I'V_{\ell}(\alpha_I)$ respectively.
It also follows that $g_Ig_I'V_{k+\ell-1}(\alpha_I)$ generates $C_ID_I$ 
(as $k+\ell -1\leq |I|$), $\dim C_ID_I=k+\ell -1=\dim CD$ and there is a unique generator matrix $G_{CD}$ of $CD$ whose $I$-indexed columns form 
$g_Ig_I'V_{k+\ell-1}(\alpha_I)$.

Let $x_0,\ldots x_{k-1}$ and $y_0,\ldots ,y_{\ell-1}$ denote the rows
of $G_C$ and $G_D$ respectively.

The key observation is the following: 
let $u,v,s,t$ be integers, $0\leq u,s\leq k-1 \quad 0\leq v,t\leq\ell-1$, such
that 
$$u+v = s+t.$$
Since $x_uy_v$ and $x_sy_t$ coincide in the $I$-indexed coordinates, and
$\dim C_ID_I = \dim CD$, the vectors $x_uy_v$ and $x_sy_t$ must coincide in every
coordinate of $\{1,\ldots ,n\}$.
In other words, if $\pi=(\pi_0,\pi_1,\dots,\pi_{k-1})^T$ and
$\rho=(\rho_0,\rho_1\dots,\rho_{\ell-1})^T$ are the $j$-th column
of $G_C$ and $G_D$ respectively, for some $j\not\in I$, then
$$
\pi_{u}\rho_{v}=\pi_{s}\rho_{t}.
$$

We now exploit this property in order to prove the lemma. Pick two columns $\pi,\rho$ of $G_C,G_D$ as above.

First assume that $\pi_0\neq 0$ and $\rho_0\neq 0$. Without loss of generality
we may assume $\pi_0=\rho_0=1$. It follows from $\pi_0\rho_1=\pi_1\rho_0$ that $\rho_1=\pi_1=:\beta\in F$.
For all $i\leq k-1$, it holds that $\pi_i=\pi_i\rho_0=\pi_{i-1}\rho_1$. Applying this formula recursively we obtain $\pi_i=\beta^i$ for all $i\leq k-1$, i.e.\ $\pi$ corresponds to the evaluation point $\beta\in F$. The same argument applies to $\rho$, which corresponds to the evaluation point $\beta\in F$ as well.


Now assume that $\pi_0=0$.
If $\rho_0\neq 0$, then $\pi_1\rho_0=\pi_0\rho_1=0$
implies $\pi_1=0$. Continuing in this way, we see that
if $\pi_i=0$, then $\pi_{i+1}\rho_0=\pi_i\rho_1=0$
implies $\pi_{i+1}=0$ and by induction we obtain $\pi =0$ which contradicts
the full-support property of the MDS code $C$. Therefore $\rho_0=0$.
Assume without loss of generality that $k\leq \ell$.
If $k=\ell=2$, then both $\pi_1$ and $\rho_1$ are non zero as $C$ and $D$ have full support, hence the columns $\pi$ and $\rho$ correspond to the evaluation point $\infty$.
If $k=2$ and $\ell \geq 3$ then as $\rho_i\pi_1=\rho_{i+1}\pi_0=0$ for all $i<\ell-1$ and as $\pi_1\not=0$ it follows that $\rho_i=0$ for all $i<\ell-1$ and again the full-support property of $D$ implies that the column $\rho$ corresponds to the evaluation point $\infty$.
%
If $k>2$, then the 
same procedure that we  applied to $\pi_0,\rho_0$ again
yields $\pi_1=\rho_1=0$.
Iterating in this way, 
we obtain that both $\pi$ and $\rho$ correspond to the evaluation point $\infty$.

We have proved that up to multiplication by vectors $g,g'$, the codes $C$
and $D$ have generator matrices of the form $V_k(\alpha)$ and $V_{\ell}(\alpha)$. Since  $C$ and $D$ are MDS, 
the evaluation-sequence $\alpha$
must have distinct entries and $C$ and $D$ are Reed-Solomon codes with the
same evaluation-point sequence.
\end{Proof}

\begin{Proof}[ of Theorem~\ref{thm:vosper}]
Set $k:=\dim C$, $\ell :=\dim D$, $k^*:=\dim CD=k+\ell-1$.
Let $C_0,D_0\subseteq F^{n_0}$ be the punctured
codes obtained by projecting $C,D$ on the first $n_0:=k^*+2$ coordinates.
As $C_0,D_0$ and $C_0D_0$ are MDS, we have $\dim C_0=\dim C$,
$\dim D_0 =\dim D$, $\dim C_0D_0 = \dim CD$ and
\begin{equation}\label{eq:projidentity}
k^*=\dim C_0D_0=\dim C_0+\dim D_0-1=n_0-2.
\end{equation}
Define the code $A\subseteq F^{n_0}$ by
$$A:=(C_0D_0)^\bot.$$
By Lemma~\ref{lemma:mds_product} the code $C_0D_0$ is MDS, therefore
$A$ is MDS and furthermore has dimension~$2$ by 
\eqref{eq:projidentity}.
Now observe that for any $a\in A$, $x\in C_0$, $y\in D_0$, orthogonality
of $A$ and $C_0D_0$ translates into
$$\scalprod{a}{xy}=0$$
which is equivalent to
$$\scalprod{ax}{y}=0.$$
We have therefore $(AC_0)^\bot\supseteq D_0$, from which we deduce
$$
\dim AC_0\leq n_0-\dim D_0=\dim C_0+1\leq n_0-1
$$
whence
\begin{equation}
  \label{eq:AC0}
  \dim AC_0 = \dim C_0 +1
\end{equation}
by Theorem~\ref{thm:mdsbound}. Similarly we also have
\begin{equation}\label{eq:AD0}
  \dim AD_0 = \dim D_0 +1.
\end{equation}

Now $A$ is an MDS code of dimension $2$ and therefore has
a generator matrix with at most two zero entries. By puncturing
one coordinate if need be, we obtain a generator matrix with
at most one zero entry. The two
rows of this matrix are clearly of the form $g,g\alpha$ for
some $g\in F^n$ and $\alpha\in F^n$ with pairwise distinct coordinates.
Finally, consider that $\dim C_0=n_0-1-\dim D_0\leq n_0-3$, and
similarly $\dim D_0 \leq n_0-3$. Hence \eqref{eq:AC0} and \eqref{eq:AD0}
imply 
\begin{align*}
  \dim AC_0 &\leq n_0-2,\\
  \dim AD_0 &\leq n_0-2.
\end{align*}
Therefore Lemma~\ref{lemma:lemma4} applies to $A,C_0$ and to $A,D_0$,
possibly after puncturing one coordinate. From there we obtain that
$C_0$ and $D_0$ (possibly punctured on a common coordinate) are
Reed-Solomon codes with a common evaluation-point sequence, and
Lemma~\ref{lemma:columns} gives the desired conclusion.
\end{Proof}

An interesting consequence of Theorem~\ref{thm:vosper} is the following
characterization of Reed-Solomon codes among MDS codes. Applying
Theorem~\ref{thm:vosper} in the case $C=D$ yields:

\begin{Corollary}\label{cor:test}
Let $C\subseteq F^n$ be an MDS code, with $\dim C\leq (n-1)/2$. 
The code $C$ is Reed-Solomon if and only if
\begin{equation}\label{eq:square}
   \dim C^2=2\dim C-1.
\end{equation}
\end{Corollary}

\begin{Remark}
If $\dim C\geq (n+1)/2$, then $C$ being MDS we must have $C^2=F^n$
and the dimension of the square cannot yield any information on the
structure of $C$. However in that case, whether $C$ is Reed-Solomon is
betrayed by the dimension of the square of the dual code $C^\bot$.
The remaining case in which Corollary~\ref{cor:test} does not
say anything is the case $\dim C =n/2$.
One may wonder whether it still holds that $C$ is Reed-Solomon if and only if $\dim C^2=2\dim C-1$, and possibly Theorem~\ref{thm:vosper} and Corollary~\ref{cor:test}
have not managed to capture this fact.

The answer to this question is negative, indeed
 there exist plenty of MDS codes of dimension $n/2$
satisfying \eqref{eq:square}
 which are not Reed-Solomon.
For instance, the codes denoted $C_{11,8,8}$ and $C_{13,8,21}$
in~\cite{BGGHK03}, of length $8$ over the fields with $11$ and $13$ elements
respectively are self-dual, therefore satisfy \eqref{eq:square},
and can be shown not to be Reed-Solomon.
\end{Remark}

\section{Classification of PMDS pairs}\label{sec:main}
We now are finally ready to focus on the paper's central result, 
namely Theorem~\ref{thm:main}.

First, we show how Randriambololona's Product Singleton Bound can be obtained as a consequence of Theorem~\ref{thm:kneser}.
To be precise we obtain:


\begin{Theorem}\label{thm:psb}
Let $C_1,\dots,C_t\subseteq F^n$ be codes. Assume that their product $C_1\cdots C_t$ has full support. Then
$$
d_{\min}(C_1\cdots C_t)\leq\max\{t-1,n-(\dim C_1+\dots+\dim C_t)+t\}.
$$
\end{Theorem}

\begin{Remark}
 The full result of \cite{Randriam13_psb} is actually stronger than
 Theorem~\ref{thm:psb},
 as it ensures that an element of weight at most $\max\{t-1,n-(\dim C_1+\dots+\dim C_t)+t\}$ can be found in the set
$$
\{x_1\cdots x_t: x_1\in C_1,\dots,x_t\in C_t\},
$$
and not only in its span. 
The support condition given here is also not the same as
the apparently weaker hypothesis given in \cite{Randriam13_psb},
but the two conditions are really interchangeable, as
argued in~\cite[Remark 3(c)]{Randriam13_psb}.
\end{Remark}

\begin{Proof}[ of Theorem~\ref{thm:psb}]
For ease of notation, set $k_i:=\dim C_i$ for all $i=1,\dots,t$, $P:=C_1\cdots C_t$, $k^*:=\dim P$, $d^*:=d_{\min}(P)$. Assume that $d^*\geq t$.
The classical Singleton Bound, applied to $P$, says that
\begin{equation}\label{eq:singleton}
k^*\leq n-d^*+1.
\end{equation}
Repeatedly applying Kneser's Theorem~\ref{thm:kneser} we obtain
\begin{equation}\label{eq:t-kneser}
k^*\geq k_1+\dots+k_t-(t-1)\dim\St(P).
\end{equation}
Combining it with~\eqref{eq:singleton}, we get
\begin{equation}\label{eq:kneser}
d^*\leq n-(k_1+\dots+k_t)+1+(t-1)\dim\St(P),
\end{equation}
which is apparently a weaker statement than Theorem~\ref{thm:psb}.
To improve it, we ``correct''~\eqref{eq:singleton} to transform it into an identity, namely we define $m:=n-d^*+1-k^*$. Thus, by definition, $P$ is ``$m$-far from being MDS''. The combination of this identity with~\eqref{eq:t-kneser} gives an improved version of~\eqref{eq:kneser}, namely
\begin{equation}
d^*=n-k^*+1-m\leq n-(k_1+\dots+k_t)+1+(t-1)\dim\St(P)-m.\label{eq:psb3}
\end{equation}
In the case of $t=2$, the first claim of Lemma~\ref{lemma:stabilizer}, rewritten as
$$
\dim\St(P)-(n-d^*+1-k^*)\leq1
$$
immediately proves the theorem.
In the general case, using the second claim of Lemma~\ref{lemma:stabilizer} instead we obtain
\begin{align}
(t-1)\dim\St(P)-m&\leq (t-1)\ \frac{n-k^*}{d^*-1}-m\nonumber\\
&=t-1+(t-1)\ \frac{m}{d^*-1}-m\nonumber\\
&=t-1-\frac{d^*-t}{d^*-1}\ m.\label{eq:psb4}
\end{align}
As $d^*\geq t$ the conclusion follows.
\end{Proof}

From here on we focus on the case of $t=2$. Recall that a pair 
of codes $C,D\subseteq F^n$ is defined to be
PMDS if 
$$2\leq d_{\min}(CD)=n -\dim C -\dim D +2.$$
Observe that for a PMDS pair $(C,D)$ all inequalities in the proof of Theorem~\ref{thm:psb} are actually identities.
From this simple observation we obtain some corollaries which relate 
the Product Singleton Bound with Kneser's Theorem and with the classical Singleton Bound.

\begin{Corollary}\label{cor:pmds_properties}
Let $C,D\subseteq F^n$ be codes such that the pair $(C,D)$ is PMDS. Then the following hold.
\begin{enumerate}
\item The pair $(C,D)$ attains the bound of Kneser's Theorem, i.e.\
$$
\dim CD=\dim C+\dim D-\dim\St(CD).
$$
\item Either $CD$ is MDS or $d_{\min}(CD)=2$.
\end{enumerate}
\end{Corollary}
\begin{Proof}
From the above observation,~\eqref{eq:t-kneser} is an identity if $(C,D)$ is PMDS, hence the first claim is immediately proved.
From~\eqref{eq:psb3} and~\eqref{eq:psb4} we obtain
$$
\frac{d_{\min}(CD)-2}{d_{\min}(CD)-1}\ m=0,
$$
where $m:=n-d_{\min}(CD)+1-\dim CD$, hence either $m=0$ meaning $CD$ is MDS,
or $d_{\min}(CD)=2$.
\end{Proof}

The two possible cases in our main Theorem~\ref{thm:main} arise from the two possible situations given by the second claim of the above corollary. We distinguish the case of $d_{\min}(CD)>2$, which implies that $CD$ is MDS, and $d_{\min}(CD)=2$.

\begin{Proposition}\label{prop:case1}
Let $C,D\subseteq F^n$ be codes such that the pair $(C,D)$ is PMDS, and assume $d_{\min}(CD)>2$.
Then $C,D$ and $CD$ are MDS.
Moreover, if $\dim C,\dim D\geq2$ then $C,D$ and $CD$ are Reed-Solomon codes with a common evaluation-point sequence.
\end{Proposition}
\begin{Proof}
By the above corollary $CD$ is MDS. Moreover the PMDS property immediately yields $n>\dim C+\dim D$.
We now proceed to prove that $C$ and $D$ are also MDS through 
Lemma~\ref{lemma:mds_equivalent}.

Set $k:=\dim C,\ell:=\dim D$. Without loss of generality,
we can choose a generator matrix $G_C$ of $C$ that is
 systematic in the first $k$ positions. 
Let $G_D$ be a generator matrix of $D$.
The matrix formed by the last $n-k$ columns of $G_D$ has full rank, 
otherwise there is a non-zero vector of $D$ that is zero in the last
$n-k$ positions, and taking the product with a row of $G_C$ we
would obtain a vector of $CD$ of weight $1$, contradicting that $CD$ is
MDS and not the whole space $F^n$.
So we can now assume that $G_C$ is systematic in the first $k$ positions and $G_D$ is systematic in the subsequent $\ell$ positions.

Now we focus on $G_C$.
Assume that there is a zero entry in the $j$-th column of $G_C$ for some
$j>k+\ell$, say in position $(i,j)$ of $G_C$. Then, since
the $j$-th column of $G_D$ is not all-zero (otherwise $CD$ would not be full
support and would not be MDS), the product of the $i$-th row of $G_C$
with some row of $G_D$ yields non-zero vector of $CD$ of 
weight at most $n-k-\ell+1=d_{\min}(CD)-1$, a contradiction. 
Therefore, all columns of $G_C$ indexed by $j>k+\ell$, that
exist since $n>k+\ell$, have no zero entries. For the same reason,
this is also true of $G_D$, and we obtain that the product of
any row of $G_C$ with any row of $G_D$ is non-zero.

From this last fact, we get that
$G_C$ cannot have zero entries in the columns indexed by $\{k+1,\dots,k+\ell\}$, or again, by taking a product of a row of $G_C$ with a row of 
$G_D$, we would have a non-zero vector
 of $CD$ of weight at most $n-k-\ell+1$.
Now Lemma~\ref{lemma:mds_equivalent} allows us to conclude that $C$ is MDS.
Analogously, one has that $D$ is MDS as well.

The last statement now follows immediately by Theorem~\ref{thm:vosper}.
Note that $n>\dim C+\dim D$ is equivalent to the hypothesis $\dim CD\leq n-2$.
\end{Proof}

The following lemma will be useful to deal with the second case.

\begin{Lemma}\label{lemma:conclusion}
Let $C,D\subseteq F^n$ be codes such that $CD$ is MDS and
$$
\dim CD=\dim C+\dim D-1=n-1.
$$
Then there exists $g\in (F^n)^\times$ such that $C=(gD)^\bot$.
\end{Lemma}
\begin{Proof}
Let $g\in F^n$ be a generator of $(CD)^\bot$, which is invertible 
as $(CD)^\bot$ is MDS of dimension $1$. 
For any $x\in C$, $y\in D$, we have
$$
\scalprod{x}{gy}=\scalprod{xy}{g}=0
$$
so that $C\subseteq (gD)^\bot$, and equality follows by comparing dimensions.
\end{Proof}

\begin{Proposition}\label{prop:case2}
Let $C,D\subseteq F^n$ be codes such that the pair $(C,D)$ is PMDS.
Set $h:=\dim\St(CD)$ and let $\{\pi_1,\dots,\pi_h\}$ be an $F$-basis of $\St(CD)$ of disjoint projectors with supports $I_1,\dots,I_h$.
Then $C,D$ and $CD$ decompose as
\begin{align*}
C&=\pi_1C\oplus\dots\oplus \pi_hC,\\ D&=\pi_1D\oplus\dots\oplus \pi_hD,\\ CD&=\pi_1CD\oplus\dots\oplus \pi_hCD
\end{align*}
and, for all $i=1,\dots,h$, we have $\supp\pi_iC=\supp\pi_iD=\supp\pi_iCD=I_i$ and
\begin{equation}\label{eq:reverse1}
\dim \pi_iCD=\dim \pi_iC+\dim \pi_iD-1.
\end{equation}
Moreover, if $d_{\min}(CD)=2$ then, for all $i=1,\dots,h$, 
when $\pi_iC$ and $\pi_iD$ are identified with codes of $F^{|I_i|}$ through the natural projection on their support, then $\pi_iC=(g_i\pi_iD)^\bot$ for some $g_i\in (F^{|I_i|})^\times$.
\end{Proposition}
\begin{Proof}
By Kneser's Theorem we have, for all $i=1,\dots,h$,
\begin{equation}
  \label{eq:pi_i1}
  \dim \pi_iCD\geq\dim \pi_iC+\dim \pi_iD-1
\end{equation}
since $\pi_iCD$ has trivial stabilizer. Therefore
\begin{equation}
  \label{eq:pi_i2}
 \dim CD=\sum_{i=1}^h\dim \pi_iCD\geq \sum_{i=1}^h(\dim \pi_iC+\dim \pi_iD-1).
\end{equation}
Observing that
\begin{equation}\label{eq:inclusion}
  C\subseteq\pi_1C\oplus\cdots\oplus\pi_hC,\quad
  D\subseteq\pi_1D\oplus\cdots\oplus\pi_hD
\end{equation}
we get
  $$\sum_{i=1}^h(\dim \pi_iC+\dim \pi_iD-1)\geq \dim C+\dim D-h,$$
but the right hand side of this inequality equals $\dim CD$
by the first claim of Corollary~\ref{cor:pmds_properties}. From
\eqref{eq:pi_i2} we obtain therefore that all inequalities in
\eqref{eq:pi_i1} are equalities, i.e.\ for all $i=1,\dots,h$, 
$$
\dim \pi_iCD=\dim \pi_iC+\dim \pi_iD-1,
$$
and we obtain also
$$
\sum_{i=1}^h(\dim \pi_iC+\dim \pi_iD) = \dim C + \dim D,
$$
hence both inclusions in~\eqref{eq:inclusion} are actually identities.

Now assume that $d_{\min}(CD)=2$. Observe that $n=\dim C+\dim D$ by the Product Singleton Bound.
From here on all codes are identified with full-support codes through the natural projection on their support.
For all $i=1,\dots,h$, we have $d_{\min}(\pi_iCD)\geq2$, hence $|I_i|\geq\dim \pi_iCD+1$ by the classical Singleton Bound applied to $\pi_iCD$.
Therefore
\begin{align*}
n=\sum_{i=1}^h|I_i|&\geq \sum_{i=1}^h(\dim \pi_iCD+1)\\&=\sum_{i=1}^h(\dim \pi_iC+\dim \pi_iD)\\&=\dim C+\dim D=n.
\end{align*}
It follows that
$$
\dim\pi_iCD=\dim \pi_iC+\dim \pi_iD-1=|I_i|-1
$$
and $d_{\min}(\pi_iCD)\geq2=|I_i|-\dim\pi_iCD+1$ proves that $\pi_iCD$ is MDS. Now the conclusion follows by Lemma~\ref{lemma:conclusion}.
\end{Proof}

Propositions~\ref{prop:case1} and~\ref{prop:case2} constitute the proof of Theorem~\ref{thm:main}.

\section{Concluding comments}\label{sec:comments}
As mentioned in Section~\ref{sec:vosper}, 
Theorem~\ref{thm:vosper} is arguably a coding-theoretic analogue
of Vosper's Addition Theorem. The analogy with its additive counterpart
is not as clear-cut however as in the case of Theorem~\ref{thm:kneser}
and Kneser's Addition Theorem. More precisely,
the MDS hypothesis in Theorem~\ref{thm:vosper}
is not a very natural analogue of the prime order of the ambient group
hypothesis in Vosper's original Theorem, and there may possibly be
other coding-theoretic analogues to consider.

The natural question raised by Theorem~\ref{thm:kneser}
and Theorem~\ref{thm:vosper} is whether there
exists a satisfying characterization of pairs $C,D$ such that
$CD$ is indecomposable and of codimension at least $2$,
and $\dim CD =\dim C + \dim D -1.$
Beside pairs of Reed-Solomon codes, one now has
Reed-Solomon codes with duplicate coordinates.
Beside these, other examples turn up. In particular,
one may take the amalgamated direct sum \cite{CHLL97}
of self-dual codes.

If the analogy with additive combinatorics is to be trusted,
such a characterization may be tractable -- though probably
difficult -- and would be a coding-theory equivalent of
Kemperman's Structure Theorem for small sumsets~\cite{Kemperman}.

Finally, it is natural to wonder whether the characterization of PMDS pairs extends to products of more than two codes.
Our techniques (Corollary~\ref{cor:pmds_properties} and Proposition~\ref{prop:case1}) allow to deal with the analogue of the first case of Theorem~\ref{thm:main} and to prove the following: 
if $(C_1,\dots,C_t)$ is a $t$-PMDS tuple, i.e.\ satisfies
$$
d_{\min}(C_1\cdots C_t)= n-(\dim C_1+\dots+\dim C_t)+t,
$$
if none of the $C_i$'s has dimension $1$ and
$$
d_{\min}(C_1\cdots C_t)>t,
$$
then all $C_i$'s are Reed-Solomon codes with a common evaluation-point sequence.
On the other hand the arguments in the paper do not seem quite sufficient to deal with the case of
$$
d_{\min}(C_1\cdots C_t)=t
$$
corresponding to the second case of our main theorem.
We leave the matter open for further study.

While this paper was in review, a version of Kneser's Theorem for a family of algebras was independently posted by Beck and Lecouvey~\cite{BL15}, and is a more general version of Theorem~\ref{thm:kneser}. The proof follows similar arguments.
Moreover, ~\cite[Section~4.3]{BL15} shows how Kneser's original Theorem can be recovered from the new variant. The very same argument, which is based on the embedding of the group $G$ into the complex group algebra $\C[G]$ and on the isomorphism $\C[G]\cong\C^{|G|}$, allows one to recover the original theorem from our variant as well.

\end{document}